\documentclass[10pt]{article}   
\pdfoutput=1
\usepackage{jheppub}
\usepackage{amsmath,amssymb,amsfonts,mathbbol,graphicx,slashed,color,amsthm, mathtools, upgreek, enumerate, tensor}

\usepackage{subcaption}
\usepackage{setspace}
\usepackage[export]{adjustbox}
\usepackage{arydshln}
\usepackage[dvipsnames]{xcolor}
\usepackage{physics}
\usepackage{hyperref}
\usepackage{comment}
\graphicspath{{pics/}}
\usepackage{breqn}

\usepackage{comment}

\usepackage{algorithm}
\usepackage{algpseudocode}

\allowdisplaybreaks

\colorlet{darkblue}{blue!70!black}

\colorlet{darkgreen}{green!50!black}
\colorlet{darkred}{red!50!black}

\newcommand{\mc}{\mathcal}

\renewcommand{\t}{\tilde}

\def\bea{\begin{eqnarray}}
\def\eea{\end{eqnarray}}
\def\be{\begin{equation}}
\def\ee{\end{equation}}

\title{
 Timelike-bounded $dS_4$ holography from a solvable sector of the $T^2$ deformation
}

\author[a]{Eva Silverstein,}
\author[b]{Gonzalo Torroba}
\affiliation[a]{Stanford Institute for Theoretical Physics, 382 Via Pueblo, Stanford, CA 94305}
\affiliation[b]{Centro At\'omico Bariloche, CONICET, and Instituto Balseiro, Bariloche, RN, Argentina}
 
 \vspace{5mm}

\vspace{1cm}

\abstract{

Recent research has leveraged the tractability of $T\bar T$ style deformations to formulate timelike-bounded patches of three-dimensional bulk spacetimes including $dS_3$.  This proceeds by breaking the problem into two parts:  a solvable theory that captures the most entropic energy bands, and a tuning algorithm to treat additional effects and fine structure.  We point out that the method extends readily to higher dimensions, and in particular does not require factorization of the full $T^2$ operator (the higher dimensional analogue of $T\bar T$ defined in \cite{Hartman:2018tkw}).    Focusing on $dS_4$, we first define a solvable theory at finite $N$ via a restricted $T^2$ deformation of the $CFT_3$ on ${S}^2\times \mathbb{R}$, in which $T$ is replaced by the form it would take in symmetric homogeneous states, containing only diagonal energy density $E/V$ and pressure (-$dE/dV$) components.  This explicitly defines a finite-N solvable sector of $dS_4/\text{deformed-CFT}_3$, capturing the radial geometry and count of the entropically dominant energy band, reproducing the Gibbons-Hawking entropy as a state count.  To accurately capture local bulk excitations of $dS_4$ including gravitons, we build a deformation algorithm in direct analogy to the case of $dS_3$ with bulk matter recently proposed in \cite{Batra:2024kjl}. This starts with an infinitesimal stint of the solvable deformation as a regulator. The full microscopic theory is built by adding renormalized versions of $T^2$ and other operators at each step, defined by matching to bulk local calculations when they apply, including an uplift from $AdS_4/CFT_3$ to $dS_4$ (as is available in hyperbolic compactifications of M theory).   The details of the bulk-local algorithm depend on the choice of boundary conditions; we summarize the status of these in GR and beyond, illustrating our method for the case of the cylindrical Dirichlet condition which can be UV completed by our finite quantum theory.
 }

\begin{document}

\maketitle
\parskip=10pt

\section{Introduction}\label{sec:intro}

Holography with positive cosmological constant is a key goal of quantum gravity research.
Several developments, on timelike boundaries \cite{Anderson:2006lqb, Anderson:2010ph, Bredberg:2011xw, Anninos:2011zn, Andrade:2015gja, Marolf:2012dr, An:2021fcq, Anninos:2022ujl,  Miyashita:2021iru, Draper:2022ofa, Banihashemi:2022htw, Banihashemi:2022jys, Anninos:2023epi, Anninos:2024wpy, Liu:2024ymn, Banihashemi:2024yye}, refined entropy counts \cite{Anninos:2020hfj}, uplifts of AdS/CFT \cite{Dong:2010pm, DeLuca:2021pej}, $T\bar T$-style deformations \cite{Zamolodchikov:2004ce, Smirnov:2016lqw, Cavaglia:2016oda, Dubovsky:2012wk, Dubovsky:2017cnj, Dubovsky:2018bmo, Freidel:2008sh, McGough:2016lol, Kraus:2018xrn, Hartman:2018tkw, Gorbenko:2018oov, Lewkowycz:2019xse} including pioneering work by Freidel \cite{Freidel:2008sh}, and the role of multitrace deformations of AdS/CFT \cite{Aharony:2001dp, Aharony:2001pa} as boundary conditions \cite{Berkooz:2002ug, Witten:2001ua}, among others,  have recently come together to help formulate de Sitter \cite{Galante:2023uyf}.  Timelike boundaries play a key role, introducing a well defined boundary energy spectrum consistent with the constraints of general relativity. This approach has led so far to a prescription for generating a finite quantum Hamiltonian formulating quantum gravity with positive cosmological constant in three bulk dimensions, whose spectrum produces a well-defined cosmic horizon microstate count \cite{Coleman:2021nor, Batra:2024kjl}.
\footnote{This development enjoys further synergy with other recent approaches to local de Sitter holography such as \cite{Chandrasekaran:2022cip, Narovlansky:2023lfz, Susskind:2021dfc, Rahman:2024vyg} as well as earlier works such as \cite{Anninos:2011af, Banks:2006rx, Bredberg:2011xw, Anninos:2011zn, Donnelly:2016auv, Ciambelli:2023mir, Ciambelli:2024swv} .}  

This is in accord with recent arguments that there is only one state of {\it closed} universes in quantum gravity (see e.g. \cite{Usatyuk:2024isz}). If that is the case, then given the observed positive cosmological constant, any microstate count in quantum gravity -- whether for black holes or the cosmic horizon -- requires some other topology, such as the timelike boundaries at play here.  Such boundaries, when they are consistent (see section \ref{subsec:timelike-boundaries}), introduce a nontrivial energy spectrum to count. Moreover, the presence of one or more such boundaries is generic -- global dS, where gravity does not decouple, is a very special case.

The results \cite{Coleman:2021nor, Batra:2024kjl} were specific to a 3-dimensional bulk theory.
In this note, we refine the method and point out that it readily generalizes to the case of four bulk dimensions, even though the special properties of the $T\bar T$ operator in two dimensions \cite{Zamolodchikov:2004ce, Smirnov:2016lqw} do not extend to the full $T^2$ operator in higher dimensions \cite{Hartman:2018tkw, Taylor:2018xcy}.   For boundary dimension $d=3$ there is still a sector -- containing the most entropic bands of energy levels  -- that is captured by a solvable deformation of the $AdS_4/CFT_3$ Hamiltonian similar to \cite{Coleman:2021nor}.  
We define this solvable theory by restricting the $T^2$ deformation of AdS/CFT (as derived at large $N$ in \cite{Hartman:2018tkw} and generalized here to include a constant $\Lambda_3$ addition at each step) to 
contain only diagonal 
components of the stress tensor and to take the form that that it does in symmetric homogeneous states,
depending only on energy density $E/V$ and pressure $-dE/dV$.  
The finite real spectrum of this Hamiltonian, manifest in equation \eqref{eq-4d-dressed-energies} below, captures the Gibbons-Hawking entropy along with the radial geometry of the $dS_4$ cosmic horizon patch as a function of boundary size. The operator algebra, working at finite $N$, is type I, simpler than the type II result one obtains in the $G_N\to 0$ limit \cite{Chandrasekaran:2022cip}.

\begin{figure}[h]   
\begin{center}
  \includegraphics[width=0.8\textwidth]{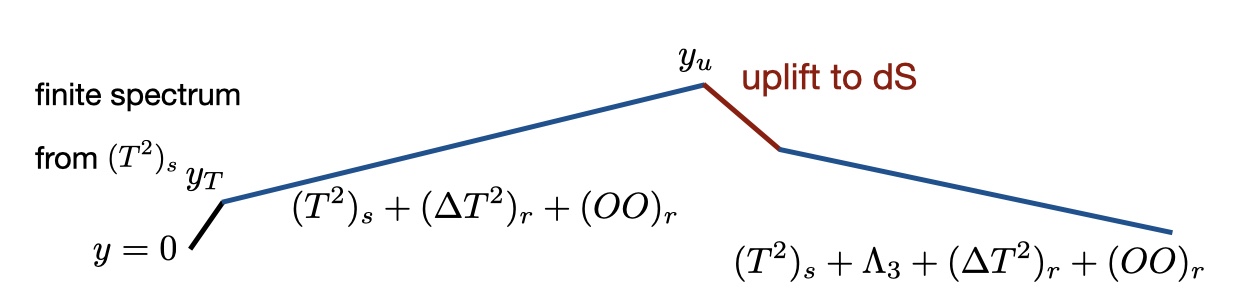}
      \caption{Schematic representation of the algorithm proposed in this work, adapted from \cite{Batra:2024kjl}. An infinitesimal application of the solvable deformation $(T^2)_s$ defined in \S\ref{sec-solvable-deformation} up to deformation parameter $y_T$ produces a unitary quantum theory with a finite real spectrum. Then we proceed with a step by step application of the renormalized $T^2$ operator $(T^2)_r = (T^2)_s+ (\Delta T^2)_r$, along with renormalized matter contributions.  These operators are defined (see \S\ref{sec-local-algorithm}) so that their matrix elements (including correlators) match those of the bulk theory when those are available, including sectors which will be involved in the transition from AdS to dS. This uplift  \cite{DeLuca:2021pej} is done at a value of the deformation parameter $y=y_u$ where for the most entropic band of energies the square root in \eqref{eq-4d-dressed-energies1} vanishes.  We then proceed with the renormalized $(T^2)_r+\Lambda_3$ deformation with matter.  The result is a timelike boundary theory whose spectrum includes the timelike-bounded $dS_4$ cosmic horizon patch, pole patch, and local bulk excitations thereof (see Fig. \ref{fig:states}).  The $(\Delta T^2)_r + (OO)_r$ contributions are subleading in the entropy, which along with the radial geometry is captured by the solvable version of the theory $(T^2)_s + \Lambda_3$ defined in \S\ref{sec-solvable-deformation}.}
    \label{fig:setup}
\end{center}
\end{figure}

\begin{figure}[h]   
\begin{center}
  \includegraphics[width=0.6\textwidth]{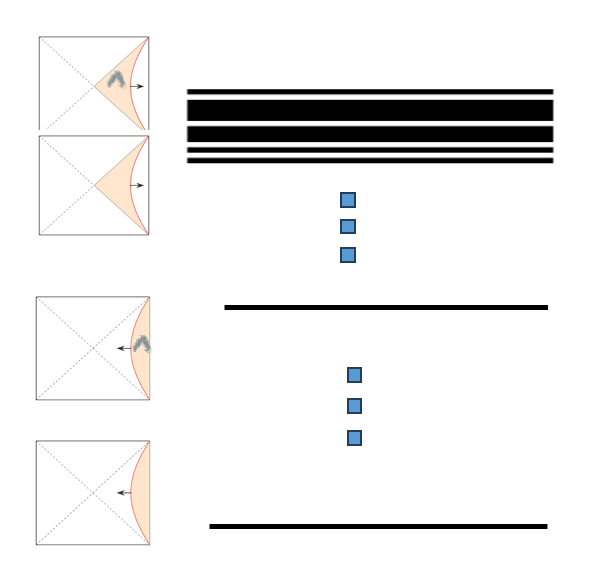}
      \caption{The result of our procedure is quantum mechanics theory whose spectrum includes bands of energy levels dual to the timelike-bounded cosmic horizon patch, pole patch, and local excitations therein.  The arrows depict the branch choice of the square root in the energy formula \eqref{eq-4d-dressed-energies2}; both choices appear in the spectrum of the full theory as explained in \cite{Batra:2024kjl}.}
    \label{fig:states}
\end{center}
\end{figure}

This Hamiltonian does not accurately capture the local dynamics of additional excitations including the bulk gravitons which are a new feature in $d+1=4$ dimensional bulk gravity.  
The gravitons and other entropically subdominant bulk-$4d$ excitations are analogous to matter field excitations in bulk-$3d$ effective field theory.  Similarly to the treatment of matter in $3d$ in \cite{Batra:2024kjl}, we can embed the solvable sector into a more general theory -- defined by a step by step algorithm updating the Hamiltonian along the trajectory -- to capture these excitations and their boundary conditions accurately.  This begins with an infinitesimal stint of the solvable deformation to produce a starting point with a finite spectrum.  Due to the vast state space of quantum systems, there is plenty of capacity in this finite theory to define the appropriate multitrace operators needed to deform the boundary conditions of bulk fields, ensuring that they match the relatively small set of observables captured by controlled calculations in the low energy theory. We will include in the set of bulk quantities to match some aspects of M theory involved in the uplift, including bulk degrees of freedom that become light in either the bounded dS or bounded AdS phase of our deformation.  We stress that the entire effective potential, including localized topology change processes required to uplift from $S^7$ to the negatively curved compactification \cite{DeLuca:2021pej}, is sub-Planckian in energy density, as can be seen by expressing the effective potential in Einstein frame \cite{Silverstein:2004id, Douglas:2009zn}.  This procedure is summarized in Fig. \ref{fig:setup}, leading to a quantum theory whose spectrum contains bands of energies corresponding to bulk gravitational solutions including the $dS_4$ cosmic horizon patch, the pole patch, and local bulk excitations of these geometries as depicted in Fig. \ref{fig:states}.


The procedure we develop here applies to a boundary theory with $d \ge 2$ space-time dimensions, but $T\bar T$ methods originally developed for $d=2$ enter in two essential ways.  First, the notion of defining a theory via a deformation, which we can translate into a step by step algorithm, is extremely useful more generally, beyond the special case of $d=2$ $T\bar T$.  Secondly, as we will see in detail below, our solvable Hamiltonian for boundary dimension $d=3$ satisfies a differential equation similar to the full equation for the energies in $d=2$ \cite{Zamolodchikov:2004ce, Smirnov:2016lqw, Coleman:2021nor}.

The general strategy is to define a theory which, by construction, matches the observables of the bulk effective theory when they apply, while filling in microscopic details which are a priori unknown. Key among these is the microstate count for the bounded $dS_4$ cosmic horizon patch, which is conveniently captured by the solvable sector of the theory.  Aside from generalizing the large-N treatment of \cite{Hartman:2018tkw} -- including its calculations of homogenous energy levels -- to include the $\Lambda_3$ contribution, the contribution of this paper is largely conceptual, pointing out how the methods developed in \cite{Batra:2024kjl, Coleman:2021nor} readily generalize to four dimensions.  Along the way, we will generalize the treatment in \cite{Batra:2024kjl} in several ways, and will find additional uses of the boundary theory for understanding the consistency of timelike boundaries.       

\subsection{Status of classical and quantum timelike boundaries}\label{subsec:timelike-boundaries}

Before continuing, we should address the fact that classifying consistent boundary conditions at timelike boundaries at the level of general relativity, effective field theory, string theory, and holography, is an ongoing research direction. This is a rich problem with many facets (for early reviews see \cite{Szabados:2004xxa, Sarbach:2012pr}) that we attempt to briefly summarize here, with an eye toward the focus of the present work.  In the bulk of this paper, we will specialize to a viable special case of Dirichlet boundary conditions, exploiting the structure of our boundary theory to help define it.

At the level of general relativity, the objective is to construct or rule out classes or particular examples of boundary conditions satisfying the criterion that there exist unique solutions evolved from initial data, satisfying the prescribed boundary conditions. A consistent initial boundary value problem (IBVP) has been established, for example, for totally geodesic boundaries in \cite{Fournodavlos:2020wde}, and its generalization \cite{Fournodavlos:2021eye}. Given a consistent solution to the initial boundary value problem, additional considerations include an appropriate stability analysis,  complete with an understanding of the fate and interpretation of potential instabilities.   Although we will focus here on a special case of Dirichlet boundary conditions which remains viable as described shortly, we anticipate interesting generalizations of our approach to other classes of boundary conditions. An important example is the case of conformal boundary conditions \cite{Anderson:2006lqb, Witten:2018lgb, Anninos:2024wpy, Anninos:2023epi} and generalizations \cite{Liu:2024ymn}.

The ultimate criterion is consistency at the level of the quantum theory.  This will be relevant in our results in the present work, with regard to both consistency and stability.  The importance of quantum mechanics is of course standard in physics; the stability of ordinary matter depends on it as atoms would classically decay by radiation, but in the quantum theory they persist in their stationary ground state.  Quantum effects also play an essential role in stabilizing broad classes of de Sitter models in string/M theory (see \cite{Flauger:2022hie}
for a review), and can materially affect key dynamical properties of spacetime \cite{Gao_2017, Batra:2024qju}.  Even in the presence of negative modes, quantum systems living on an unstable potential need not be physically sick, but can be defined via scattering states. Finally, and anticipating a little, in addition to stability considerations just noted, the quantum theory may address consistency issues.
In particular, in the present work, we will exploit the non-degeneracy of generic quantum spectra to avoid non-uniqueness issues in a new way.        

From the classical, low energy perspective it is well known that the initial boundary value problem is not well defined for broad classes of Dirichlet boundary conditions, at least linearized around simple backgrounds.   For example, the radial constraint linearized around these backgrounds cannot be satisfied for an {\it arbitrary} fixed boundary geometry \cite{An:2021fcq}.  This has been straightforwardly proved for perturbations of a flat boundary in Minkowski spacetime (which we will refer to as a `flat corner') to nearby boundary metrics with nonzero intrinsic scalar curvature $\delta{\mathcal R}$: with vanishing unperturbed extrinsic curvature, the linearized perturbation to the radial constraint simply reads $\delta{\mathcal {R}}=0$.  (This argument does not exclude solutions of the radial constraint with nontrivial extrinsic curvature scaling like $K\sim \sqrt{\delta g}$, which exists in many cases. Still, according to an argument in terms of Sobolev spaces in \cite{Anderson:2010ph} these nonlinear solutions do not exist for generic fixed boundary metrics, as was seen in early works on near horizon physics \cite{Bredberg:2011xw, Anninos:2011zn}.) 
At least for flat corners at the linearized level there is also a violation of geometric uniqueness \cite{Anninos:2022ujl, An:2021fcq}:  given initial data, it does not evolve to a unique solution at later times. At least for the Euclidean case, boundaries with zero extrinsic curvature also lead to problems defining a propagator for perturbation theory \cite{Anderson:2006lqb, Witten:2018lgb}.  

However,  in the special case of interest here, a cylindrical $S^2\times \mathbb{R}$ boundary geometry, the known inconsistencies that plague flat Dirichlet walls at the classical level do not apply.  In section \S5.5 of \cite{Anninos:2023epi}, the authors show in harmonic gauge that the corner analysis \cite{An:2021fcq, Anninos:2022ujl} showing a failure of geometric uniqueness in the case of flat corners does not extend to the case of a spatially spherical wall, and  \cite{Anderson:2006lqb, Witten:2018lgb} observes that a fixed sign of the trace of the extrinsic curvature mitigates problems defining perturbation theory in the Euclidean setting\footnote{when these problems do occur, since they involve short modes they are anyway UV sensitive and require a more complete treatment}. 

Next we consider the status of dynamical instabilities at the classical level. These can arise at long wavelengths amenable to a general relativity (GR) plus effective field theory (EFT) analysis, or at short wavelengths indicating UV sensitivity.   Low-energy linearized modes with a Hubble scale growth, albeit with unbounded proper oscillation near the cosmic horizon (indicating possible UV sensitivity) were observed in \cite{Andrade:2015gja}, along with growing modes in Minkowski spacetime with a timescale of order the size of the boundary.   These fluctuations deserve further study globally and nonlinearly to understand their implications.\footnote{ Inflationary mode creation is also a Hubble-scale phenomenon which does not destroy the ambient spacetime, dubbed `pseudotachyonic' in \cite{Aharony:2006ra}. 
In work in progress with Albert Law and others, we aim to investigate this with regard to both stability and potential observational signatures of boundary topology in the universe. } 
In the quantum theory of interest, such instabilities correspond to scattering states off of a potential barrier, and may fit into a well defined spectrum. 

In any case, the boundary is a short-scale feature which indicates UV sensitivity, adding to the motivation for a string/M theoretic treatment \cite{Silverstein:2022dfj, Ahmadain:2024hgd} or a boundary theory formulation.  
As noted in three bulk dimensions in \cite{Batra:2024kjl} (where no bulk gravitons exist but boundary gravitons do), the deformed theory with finite real spectrum explicitly cuts off any divergent buildup of short modes.  
In this work, we will build on this and argue that the quantum theory can help more generally to avoid consistency problems.  In particular, for a non-degenerate boundary theory, there is no room for non-uniqueness in the quantum gravity theory.  The reason is simply that for a non-degenerate spectrum, the quantum state at a given energy is uniquely fixed.  

In what follows, we will illustrate our method for the case of Dirichlet conditions on a cylinder. This builds from the treatment at the large-N level in cutoff AdS/CFT in \cite{Hartman:2018tkw} to formulate an algorithm to define a finite boundary theory by deformation from AdS/CFT.  We expect that suitable modifications of the algorithm could be built to adapt it to other possible boundary conditions.

\section{A solvable $T^2 + \Lambda_3$ deformation at finite N}\label{sec-solvable-deformation}

In this section, we will define a solvable, finite N deformation of $AdS_4/CFT_3$ which captures the microstates and radial geometry of the $dS_4$ bounded static patch in the absence of graviton or additional matter excitations.  This is in direct analogy with \cite{Coleman:2021nor}.  Then in the remainder of the paper, following the $AdS_3/CFT_2$ treatment with matter in \cite{Batra:2024kjl}, we will use this deformation as a first step to regulate the theory, defining a finite quantum mechanical system.  To capture the full set of 4d Einstein gravity states, matching all GR + EFT gravity-side phenomena (along with pertinent M theoretic quantities associated with the uplift \cite{DeLuca:2021pej}) while defining the quantum gravity theory beyond that approximation, we can build a suitable algorithm analogous to that in \cite{Batra:2024kjl}.

The works \cite{Zamolodchikov:2004ce, Smirnov:2016lqw} pioneered a method to solve for the $T\bar T$-deformed energy levels of a 2d theory, making reference to the special factorization properties of the $T\bar T$ operator that they discovered to hold in that dimensionality.  Specifically, one can define the $T\bar T = \frac{1}{8}(T_{\mu\nu}T^{\mu\nu}-(T^\mu_\mu)^2)$ operator as the surviving non-derivative term in the OPE as the constituent operators come together.  In a step by step procedure defined by a differential equation, at each step we add this spatially-integrated operator to the Hamiltonian along with a constant term dubbed $\Lambda_2$ \cite{Gorbenko:2018oov} which preserves solvability on the cylinder \cite{Lewkowycz:2019xse, Coleman:2021nor}
\begin{equation}\label{eq-Ham-def-2d}
    \lambda\partial_\lambda {H}= \frac{1}{2}\int dx\, T^\mu_\mu = \frac{1}{2}\int dx \left\{\frac{\pi \lambda}{2}(T_{\mu\nu}T^{\mu\nu}-(T^\mu_\mu)^2)+\frac{\eta-1}{\pi\lambda}\right\} \,.
\end{equation}
In writing the first equality here, we use the fact that the seed theory is a CFT and the only dimensionful parameter is $\lambda$.\footnote{One could consider more general seed theories, as in \cite{Smirnov:2016lqw}, dressing them with the solvable $T\bar T+\Lambda_2$ deformation.  For our particular holographic application to de Sitter holography, the starting theory is a CFT.}

Taking an expectation value in an energy eigenstate 
and substituting for the pressure component of the stress-energy tensor $T^x_x = -dE/dV$ (with $V$ denoting the volume of of the spatial circle of the cylinder) and similarly replacing the $T^x_0$ component in terms of the conserved momentum $J$, the equation 
for the deformed energy spectrum becomes a simple differential equation for $E = \langle E|H|E\rangle$.
Writing this in terms of a dimensionless deformation parameter $y=\lambda/V^2$ and solving it yields  
\begin{equation}\label{eq-2d-pure-gr-energy}
  {\cal E}_n= E_nL = \frac{1}{\pi y}\left(1\mp \sqrt{\eta +\frac{y}{y_0}(1-\eta)-2\pi y {\cal E}_n^{(0)}  + 4\pi^4 J^2 y^2\,}\right)\,,  ~~~~ (\text{pure}~ T\bar T+\Lambda_2),
\end{equation}
where ${\cal E}_n^{(0)}=2\pi (\Delta_n-c/12)$ is the dimensionless energy of the $n$th level in the seed CFT.  This reveals a finite real deformed spectrum.  

In the prescription \cite{Coleman:2021nor}, one starts with $\eta=1$, deforms until $y$ reaches $y_0=3/c\pi^2$, interpolates to $\eta=-1$ (resolved into an uplift achieved by appropriate matter \cite{Dong:2010pm} in \cite{Batra:2024kjl}), and proceeds back toward larger $y$ with the $\Lambda_2$ terms in place.   For this choice of $y_0$, the top band of energy levels corresponds to the boundary sitting at the horizon of the $\Delta = c/6$ BTZ black hole.  This maps onto the empty dS cosmic patch upon continuing with the $T\bar T + \Lambda_2$ deformation.  This by itself accounts for the refined Gibbons-Hawking entropy and radial geometry of the bounded $dS_3$ static patch as explained in \cite{Coleman:2021nor}. 
It  does not capture local bulk matter details treated by a more elaborate algorithm in \cite{Batra:2024kjl}.  

Some of the special properties of the stress-energy tensor in the 2d boundary theory \cite{Zamolodchikov:2004ce, Smirnov:2016lqw} do not extend to higher dimensions.   The stress-energy tensor contains more components which do not in general boil down to a thermodynamic pressure $-dE/dV$, energy density and conserved momenta. 
However, we can still define a deformed theory by specifying its Hamiltonian $H=\sum_E |E\rangle E \langle E|$, with the spectrum of energy eigenstates $\{E\}$ determined by an equation directly analogous to \eqref{eq-Ham-def-2d}, as follows.  

We work on a cylinder $S^2\times \mathbb{R}$
\begin{equation}\label{eq:metric1}
    ds^2 = -dt_{cyl}^2 + R^2 (d\theta^2 + \sin^2\theta \,d\phi^2)
\end{equation}
with $S^2$ volume $V=4\pi R^2$.  
We will work at finite $N$, but begin with the results of the work \cite{Hartman:2018tkw} (see also \cite{Taylor:2018xcy, Belin:2020oib}), which derived the double-trace deformations required to define local boundary conditions for higher-dimensional gravity at the large-N level.  In place of $T\bar T$ in \eqref{eq-Ham-def-2d}, we consider the quadratic combination of stress-energy tensor components in \cite{Hartman:2018tkw} (dubbed $T^2$ for the general-dimensional case) 
\be\label{eq:Hdef-main}
d \lambda \partial_\lambda \t H= \int d^{d-1}x\,\sqrt{-h} \left[ \frac{\pi \lambda}{d} :\left(\t T_{\mu\nu} \t T^{\mu\nu}- \frac{1}{d-1} (\t T^\mu_\mu)^2 \right):-\frac{d^2(d-1)}{4\pi} \frac{\eta-1}{\lambda}+ \frac{d}{4\pi} \left(\frac{C_d^2}{\lambda^{d-2}} \right)^{1/d} R^{(d)}\right]\,.
\ee
This corresponds to a boundary theory in $d$ space-time dimensions, with metric $h_{\mu\nu}$ and scalar curvature $R^{(d)}$. The deformation parameter $\lambda$ has dimension $(\text{length})^d$. $\t H= H+H_{CT}$ and $\tilde T_{\mu\nu}=T_{\mu\nu}+\t T^{CT}_{\mu\nu}$ are the Hamiltonian and the stress-energy tensor subtracting certain counterterms that appear in $d>2$ due to the curvature of the sphere; we will lay them out below in \S\ref{sec:def}. The $:\;\; :$ represents large-N normal ordering as described in \cite{Hartman:2018tkw}.  In our full prescription in \S\ref{sec-local-algorithm}, in place of $:T^2:$,  we will define a renomalized $(T^2)_r$ operator in a finite system which is valid at finite $N$. This operator $(T^2)_r$ splits into a piece which we will call $(T^2)_s$ which by itself leads to a solvable theory, and a remainder which will require regularization and renormalization.

For the solvable part analogous to the pure gravity case in $d=2$ \cite{Coleman:2021nor}, we start from \eqref{eq:Hdef-main} but substitute 
\begin{equation}\label{eq-Tab-equilibrium-solvable}
    T^0_0 = -E/V, ~~ T^a_b \to \delta^a_b (-dE/dV) \,,
\end{equation}
which will lead to a simple differential equation for the energy spectrum.  As just noted, we will denote the solvable $T^2$ deformation with the replacement (\ref{eq-Tab-equilibrium-solvable}) by $(T^2)_s$, which is not a local deformation.
The expressions \eqref{eq-Tab-equilibrium-solvable} faithfully capture only the entropically dominant bands of energy in the system, but the differential equation \eqref{eq:Hdef-main} with this substitution can be solved for all energy levels.  In this respect, it is analogous to the situation in $d=2$, where the pure $T\bar T+\Lambda_2$ theory accurately captures the most entropic energy bands, but it can be applied to all energy levels. In that case, accurately capturing general matter excitations requires an additional prescription \cite{Batra:2024kjl}.  In fact in contrast to the $d=2$ case, we do not faithfully capture angular momentum states in the solvable theory that we obtain using \eqref{eq-Tab-equilibrium-solvable}.  For the $d=3$ (and higher) case of interest in this work, we relegate to the additional algorithmic prescription in \S\ref{sec-local-algorithm} the faithful description of angular momentum states.  However, it might be possible to devise a more general solvable theory that captures faithfully the angular momentum states (rather than just the entropically dominant states). For more discussion of the possibility of extending the solvable theory to capture them see the end of \S\ref{sec:def} below.

Working in $d=3$,  defining the dimensionless deformation parameter $y=\lambda/V^{3/2}$ and fixing the integration constant and branch structure analogously to \cite{Coleman:2021nor, Batra:2024kjl},
the result for the energy spectrum of this solvable sector of our theory is of the form
\be\label{eq-4d-dressed-energies}
V^{1/2} E_n(y)={\mc E}_n(y) = \frac{3}{\pi y} \left(
1+2\pi(yC_3)^{2/3}\mp \sqrt{\eta+  4\pi(yC_3)^{2/3}-\frac{2\pi}{3}\left({\cal E}_n^{(0)}-\alpha(\eta)\right)y}
\,\right)\,,
\ee
where ${\cal E}_n^{(0)}$ are the undeformed energies, and $ \alpha(\eta)$ is a choice of integration constant that we will fix shortly.
We take this spectrum, along with its corresponding Hamiltonian $H=\sum_E | E\rangle E \langle E|$ to be the definition of our solvable theory.  
This spectrum arises from a deformation of AdS/CFT ($y=0, \eta=1$), with energy spectrum $E_n^{(0)} = {\cal E}_n^{(0)}/V^{1/2}$, with 
$C_3$ its degree of freedom count (e.g. order $N^{3/2}$ for the seed CFT obtained from $N$ M2 branes).  As we will see shortly, it is related to gravity-side variables in the seed theory by $C_3 =\frac{1}{12}\frac{\ell_{AdS}^{2}}{G_4} \gg 1$, with $\ell_{AdS}$ the AdS curvature radius and $G_4 = G_N$ the four-dimensional Newton constant.

We start with $\eta=1$, and dimensionless deformation parameter $y=0$.   With the upper (negative) sign of the square root we turn on $y$, with the choice $\alpha(\eta=1)=0$, until it reaches a value $y_u$:
\be\label{eq-4d-dressed-energies1}
V^{1/2} E_n(y)={\mc E}_n(y) = \frac{3}{\pi y} \left(
1+2\pi(yC_3)^{2/3}- \sqrt{1+  4\pi(yC_3)^{2/3}-\frac{2\pi}{3}{\cal E}_n^{(0)}y}
\,\right)\;,\;y<y_u\,\,.
\ee
We choose this value such that at $y=y_u$ the square root vanishes for a band of seed energy levels around ${\cal E}_{BH}^{(0)}$ corresponding to a large AdS black hole at or above the Hawking-Page transition, whose horizon radius coincides with that of a $dS_4$ model in the landscape (e.g. from the class in \cite{DeLuca:2021pej}). 
When the square root vanishes, it is insensitive to the branch and content of the square root, and we interpolate to
\be\label{eq-4d-dressed-energies2}
V^{1/2} E_n(y)={\mc E}_n(y) = \frac{3}{\pi y} \left(
1+2\pi(yC_3)^{2/3}+ \sqrt{\eta_u + 4\pi(yC_3)^{2/3}-\frac{2\pi}{3}\left({\cal E}_n^{(0)}-{\cal E}_{BH}^{(0)} \right)y}
\,\right)\,,\,y>y_u\,.
\ee
We interpolate from $\eta=1$ to a negative value $\eta_u =- 4\pi (y_u C_3)^{2/3}$, while choosing the integration constant $\alpha(\eta_u)$ so that the factor multiplying $y$ inside the square root vanishes when ${\cal E}= {\cal E}_{BH}$. As we will see shortly, this corresponds precisely to a $dS$ model with cosmological constant proportional to
\be
\frac{1}{\ell_{dS}^2}= - \frac{ \eta_u}{\ell_{AdS}^2}\,.
\ee
The previous literature \cite{Coleman:2021nor}\cite{Batra:2024kjl} focused for simplicity on $\eta_u = -1$, but given the fine discretuum of de Sitter models \cite{DeLuca:2021pej} we can readily generalize away from that case.
We then increase $y$ with the lower (positive) sign of the square root.  This builds a model reproducing the horizon entropy -- the dimension of the finite real Hilbert space -- and radial geometry of the bounded static patch for the $dS_4$ model. The horizon entropy includes the leading Gibbons-Hawking \cite{Gibbons:1977mu}  result $A/4g_N$ along with any universal logarithmic corrections 
$\propto\log(A/4 G_N)$ 
tied to the symmetry structure in the bounded AdS and dS geometries \cite{Anninos:2020hfj, Chandrasekaran:2022cip}. The radial geometry is inferred from the map between the deformed energy as a function of $y$ and the Brown-York energy as a function of the size of the bounding cylinder on the gravity side, for the appropriate band of energies corresponding to the gravity-side geometry.

Let us spell this out.  The relevant patches of (A)dS geometry can be written in the form
\begin{equation}\label{eq-f-metric}
    ds^2_4 = \frac{dr^2}{f(r)^2}-f(r)^2 dt^2 + r^2 d\Omega_2^2
\end{equation}
with 
\begin{equation}
    f(r)=\sqrt{1+ \frac{\eta r^2}{\ell^2_{AdS}}-\frac{2 G M}{r}} 
\end{equation}
where, as mentioned above, in the dS phase we will have $\eta=\eta_u <0$ and 
\be
\frac{1}{\ell^2_{dS}}=-\frac{\eta_u}{\ell_{AdS}^2}\,.
\ee
For the dS 
cosmic horizon patch, we have $M=0$, while in AdS ($\eta=1)$ $M$ denotes the mass of a black hole.  With a boundary of proper radius $R_c$,
the Brown-York energy density is
\begin{equation}\label{eq-BY-f}
    -T^{t}_t=\frac{1}{4\pi G_N}\left(C.T. \mp \frac{f(r)}{r} \right)\Big|_{r=R_c}\,,
\end{equation}
where `C.T.' denotes counterterm contributions.  Note that the field theory time coordinate $t_{cyl}$ of (\ref{eq:metric1}) is rescaled with respect to the time coordinate of (\ref{eq-f-metric}) by the gravitational redshift factor $t_{cyl}=f(R_c) t$. But the energies on both sides of the duality are the same.
Altogether,  the expression for the Brown-York energy in (A)dS in gravity-side variables, as a function of the proper radius $R_c$ of the bounding cylinder, is of the form:
\be\label{eq:End-rescaled}
E_n(r)=\frac{V_c}{4\pi G_N \ell_{AdS}} \left[b_{CT1}+b_{CT2} \,\frac{\ell_{AdS}^2}{2r^2}  \pm \sqrt{\eta+\frac{\ell_{AdS}^2}{r^2}+ \frac{c_1}{r^3}}\, \right]\Big|_{r=R_c}\,.
\ee
for bulk cosmological constant $-6\eta/\ell_{AdS}^2$. The gravitational origin of the counterterms proportional to $b_{CT1}$ and $b_{CT2}$ is discussed in \S \ref{sec:def}.

Let us write this in terms of boundary theory variables as defined above, using with the map
\be\label{eq:translation0}
C_3 =\frac{1}{12}\frac{\ell_{AdS}^{2}}{G_N}\;,\;y=\frac{\lambda}{V^{3/2}} = \frac{12 G_N \ell_{AdS}}{V_c^{3/2}}\,,
\ee
where $V$ is the deformed field theory cylinder volume, while $V_c$ is the volume of the bounding cylinder in the gravity side, which changes along the trajectory. Choosing counterterms $b_{CT 1,2}=1$, and fixing $c_1$ so as to match the seed energy spectrum,  produces the expression \eqref{eq-4d-dressed-energies}.\footnote{Away from the seed CFT limit $y\to 0$, the counterterms may be changed, but we keep their values at 1 for simplicity.}   In particular, in the $\eta=1$ (bounded AdS) part of the trajectory, focusing on entropically dominant (black hole) states, we have $c_1 = -2 G M$  in terms of the mass $M$ of the black hole.  In the bounded dS part of the trajectory, with $\eta = \eta_u <0$, and again focusing on the entropically dominant states corresponding to the cosmic horizon patch, we have $c_1=0$.

It may be useful to clarify that, while in $d=2$ the BH temperature measured on the AdS boundary and that associated to the static patch happen to be the same, they differ for $d>2$. However, our procedure involves matching microcanonical parameters --energies and volumes-- at the horizons, and the difference between the temperature in the asymptotic AdS region and that for timelike-bounded dS at the pole are not directly relevant, because the temperature (or `tomperature' \cite{Lin:2022nss}) evolves along our trajectory.
 This evolution of the temperatures along the trajectory will also be reflected in the matter bulk dynamics, such as the periodicities of 2-point functions. The fine-tuning procedure of Sec. \ref{sec-local-algorithm} will account for these properties.

In the gravity-side language, we start with the seed AdS/CFT system, with a cylindrical boundary of infinite proper size:  $R_c\to\infty$. The cylinder size $R_c$ is then reduced until it coincides with the horizon size of the AdS black hole of mass $M$.  This is the highest energy band which survives in the real spectrum of the theory, whose state count matches the black hole entropy.  At this radius $r=r_h$, $f(r_h)=0$ and the second square root term in the energy vanishes.  One cannot tell the difference between the black hole horizon and the dS cosmic horizon patch in the very near horizon regime\footnote{even internally, in the sense of \cite{Silverstein:2022dfj}}.  The $dS$ theory is obtained by uplifting $\eta$ from $+1$ to $\eta_u<0$ satisfying $\eta_u = - \ell^2_{AdS}/r_h^2$ and setting $c_1=0$.  The boundary size $R_c$ is then further reduced to retract the boundary from the horizon. At the matching point, the entropy $4\pi r_h^2/4G$ is the same in the two systems, and matches the state count of the finite real spectrum defined in \eqref{eq-4d-dressed-energies}.  In the de Sitter part of the trajectory, the energy spectrum \eqref{eq-4d-dressed-energies} retains the same state count.  It reproduces the Brown-York energy of the cosmic horizon patch as a function of the boundary cylinder size $R_c$ (therefore reproducing the $dS_4$ radial geometry).

To sum up, the reason to consider this particular Hamiltonian is the two essential features of de Sitter spacetime that it captures, as we will show explicitly in the following section:

(i) {\it Emergent radial de Sitter geometry}. The formula \eqref{eq-4d-dressed-energies}, with the $+$ branch of the square root and $\eta=\eta_u$, with the integration constant fixed as in \eqref{eq-4d-dressed-energies2},
matches the Brown-York (BY) quasilocal energy of the $dS_4$ cosmic horizon patch, as a function of dimensionless deformation parameter $y=\lambda/V^{3/2} \sim G_4\ell/V_c^{3/2}$.  As we vary $V_c$, the volume of the bounding cylinder in gravity variables, the BY energy -- which is determined by the extrinsic curvature of the boundary -- varies precisely according to the de Sitter geometry.  The $-$ branch of the square root $\eta=\eta_u$ captures the pole patch geometry in this same way.

and

(ii) {\it Gibbons-Hawking entropy as a microstate count}.  Away from $\lambda=0, \eta=1$, where the spectrum reverts to that of the seed $AdS_4/CFT_3$ theory, the real spectrum of this theory is finite.  
At $y=y_u$, 
the square root vanishes, enabling a continuous transition through the uplift. 
This band of energy levels dominates the entropy, and its state count matches the Gibbons-Hawking entropy of the cosmic horizon, whose area agrees with that of the 
black hole.  We keep the part of this finite spectrum which remains real throughout the deformation from $AdS_4/CFT_3$ ($\eta=1,$ $-$ branch of the square root) to $dS_4$ ($\eta=\eta_u$).  

We stress that this theory is completely well defined at finite $N$; its definition does not require the large-N approximation used in \cite{Hartman:2018tkw}
to ensure factorization of the local $T^2$ operator. Here, we simply define a finite quantum system by explicitly writing its Hamiltonian, which has the essential holographic properties (i) and (ii).

Moreover, as in \cite{Batra:2024kjl} we can then proceed to define a fine-tuning algorithm to capture the remaining details of $dS_4$ quantum gravity.\footnote{The $(T^2)_s$ analysis of the present section can be extended to all dimensionalities, but the complete treatment to come in \S\ref{sec-local-algorithm} based on \cite{Batra:2024kjl} may depend on dimensionality; the uplift from $AdS_4/CFT_3$ to $dS_4$ which involves matter fields has been concretely obtained in 4d \cite{DeLuca:2021pej} (as well as 3d \cite{Dong:2010pm}).}  This leverages the explicit uplift \cite{DeLuca:2021pej} of $AdS_4\times S^7/CFT_3$ from M2-branes in M theory to $dS_4$ as a compactification of M theory on an Anderson/Dehn-filled \cite{Anderson:2003un} hyperbolic  compactification with varying warp and conformal factors.  We will summarize that in \S\ref{sec-local-algorithm} after laying out the gravity side deformation required at the large-N level.

\section{Gravity side}\label{sec:def}

In this section, we fill in the gravitational background for the deformation defined above in \S\ref{sec-solvable-deformation} as well as laying the groundwork for the algorithmic extension to local bulk excitations in \S\ref{sec-local-algorithm}.   These details add the background motivation for the definition of the explicit finite-$N$ $T^2+\Lambda_3$ deformation of AdS/CFT that we defined above, which captures the entropically dominant sectors of $dS_4$ quantum gravity.  It will also provide the large-N approximation to the finite-N theory we will describe in \S\ref{sec-local-algorithm}, built by an algorithm defined to faithfully capture all states within the requisite level of approximation. 

We start by reviewing gravity with a timelike Dirichlet boundary and how it arises semiclassically in the bulk as a deformation of AdS/CFT. 
 This was derived for timelike-bounded AdS/CFT in \cite{Hartman:2018tkw} at the large-N level, which we extend to include the effect of an uplift to positive bulk cosmological constant at $y=y_u$ as in \cite{DeLuca:2021pej, Batra:2024kjl}.  We work at the level of classical GR in the bulk of this section, but comment on the scaling and suppression of quantum contributions at the end. 

We note as in \S\ref{subsec:timelike-boundaries} that although Dirichlet boundary conditions do not yield a valid initial boundary value problem in the {\it general case} \cite{An:2021fcq}, for a timelike cylindrical boundary with a spatial sphere ($S^2\times \mathbb R$) there are no known obstructions to its consistency. 
The non-uniqueness of solutions that had been seen for flat boundaries does not extend to the case of spherical boundaries \cite{Anninos:2023epi}.\footnote{The dynamics and stability of the system is a separate question.  Growing modes with singular initial data at the horizon appear in the linearized perturbation spectrum analyzed by \cite{Andrade:2015gja}, but smooth linearized perturbations do not grow without bound as a result of the diluting effects of the exponential expansion of de Sitter (REF our observables project).}  We will focus on this case, although a similar approach could work \cite{Coleman:2020jte} for the conformal boundary conditions \cite{Anderson:2003un, An:2021fcq, Anninos:2024wpy} or extensions \cite{Liu:2024ymn}.  

We work in Lorentzian signature, similarly to \cite{Batra:2024kjl}, in order to capture generic time-dependent dynamics in the theory. The gravity-side action including counterterms reads (working in arbitrary dimensionality for now)
\bea\label{eq-action-Lorentzian}
S &=& \frac{1}{16\pi G} \int_M d^{d+1}x\,\sqrt{-g} \left(R^{(d+1)} + \frac{d(d-1)}{\ell^2}\eta\right) \nonumber \\
&~~& +\frac{1}{8\pi G} \int_{\partial M} d^dx\,\sqrt{-h} \left(K - b_{CT1}\frac{d-1}{\ell}-\frac{1}{2(d-2)}{b_{CT2}}\, \ell R^{(d)}+ \ldots \right)  
\eea
where the $\dots $ refers to matter contributions and additional counterterms.  In the present work, we focus on $d=3$ and the novel aspects of 4d bulk gravity, including graviton excitations; matter effects enter very similarly to the bulk 3d case in \cite{Batra:2024kjl}. 
The Brown-York stress tensor \cite{Brown:1992bq, Balasubramanian:1999re}, including counterterms to cancel divergences in the seed theory limit, is given by  
\be\label{eq:Tgen1-main}
T_{\mu\nu}=-\frac{1}{8\pi G} \left(K_{\mu\nu}-K h_{\mu\nu}+ b_{CT1}\frac{d-1}{\ell} h_{\mu\nu}-\frac{b_{CT2}}{d-2} \ell G_{\mu\nu}^{(d)}+ \ldots \right)\,,
\ee
where for simplicity of writing we use notation with $\ell=\ell_{AdS}$, with the dS cosmological constant given by
\be
\frac{d(d-1)}{\ell_{dS}^2}=-d(d-1)\frac{\eta_u}{\ell^2}
\ee
as described in \S\ref{sec-solvable-deformation}.
Here $h_{\mu\nu}$ is the induced metric on the boundary, and $G_{\mu\nu}$ is its Einstein tensor.

As in \cite{McGough:2016lol, Kraus:2018xrn, Gorbenko:2018oov}, to derive the deformation we rewrite the radial Einstein equation
\be\label{eq:Eww}
\frac{1}{16\pi G} (K_{\mu\nu}^2- K^2)+\frac{1}{16\pi G} R^{(d)}+\frac{1}{16\pi G}  \frac{d(d-1)}{\ell^2}\eta=0\,,
\ee
in terms of the holographic stress tensor, producing an equation for its trace $T^\mu_\mu$ which generates the defining equation for the trajectory analogously to \eqref{eq-Ham-def-2d}.
We first work out this trace flow equation for (\ref{eq:Tgen1-main}), keeping up to the $G_{\mu\nu}$ term there, and not specializing to the cylinder. This is valid for general boundary metrics for $d=3, 4$. Rewriting the extrinsic curvature in terms of the holographic stress tensor, and then replacing into (\ref{eq:Eww}), we find
\bea\label{eq-TFE-curvatures-GR-variables}
\left(1+ \frac{\ell^2}{2(d-1)} R^{(d)} \right) T_\mu^\mu+ \frac{\ell^2}{(d-2)} G^{(d)}_{\mu\nu} T^{\mu\nu}&=& \frac{\ell^3}{16\pi G(d-2)^2}\left((R^{(d)}_{\mu\nu})^2-\frac{d}{4(d-4)} (R^{(d)})^2\right)\\
&+&4\pi G \ell \left(T_{\mu\nu}^2- \frac{1}{d-1}(T_\mu^\mu)^2 \right)+\frac{d(d-1)}{16\pi G \ell}(\eta-1)\nonumber\,.
\eea

Here we are interested in a boundary theory that lives on a cylinder $S^{d-1} \times \mathbb R$, and in this case it is possible to generalize (\ref{eq-TFE-curvatures-GR-variables}) to arbitrary $d$ as follows.
In $d$ dimensions, the counterterms include higher powers of invariant combinations of Riemann tensors that are necessary to cancel the seed CFT divergent terms in the partition function. See for instance \cite{Balasubramanian:1999re} for expressions up to $d=5$. For the specific case of the cylinder, it is possible to write an explicit counterterm subtraction that works for general $d$. 
The extrinsic curvature of a fixed radial slice of global AdS 
\be\label{eq:globalAdS}
ds^2= \frac{dr^2}{1+r^2/\ell^2}-(1+r^2/\ell^2) dt^2 + r^2 d\Omega_{d-1}^2\,,
\ee
is
\be\label{eq:KAdS}
K_{tt}^{(AdS)}= -\frac{1}{\ell}\sqrt{1+ \frac{\ell^2}{r^2}} \frac{r^2}{\ell^2}\;,\;K_{ij}^{(AdS)}=\frac{1}{\ell}\sqrt{1+ \frac{\ell^2}{r^2}} g_{ij}\,,
\ee
where $g_{ij}$ is the metric for the spatial sphere $S^{d-1}$ of radius $r$. Our choice of counterterm  subtraction will then be
\be\label{eq:Tct1}
T_{\mu\nu}=-\frac{1}{8\pi G} \left(K_{\mu\nu}-K h_{\mu\nu} \right)\Big|_{r=R_c}- T_{\mu\nu}^{CT}
\ee
with
\be\label{eq:Tct2}
T_{\mu\nu}^{CT}=- \frac{1}{8\pi G} \left(K_{\mu\nu}^{(AdS)}- K^{(AdS)} h_{\mu\nu} \right) \Big|_{1/r^{ \lceil d-2 \rceil }}\,.
\ee
Here the subscript means that the square root of (\ref{eq:KAdS}) is expanded to order $1/r^{\lceil d-2 \rceil}$. For $d=2$ this is just the constant term in the square root, while for $d=3, 4$ the expansion includes both the constant and the $1/r^2$ term from the square root. This is required in order to have a finite integrated energy for $r \to \infty$. This can be achieved by adding appropriate boundary terms to the action, built from local invariants of powers of the Riemann tensor, as in \eqref{eq-action-Lorentzian}.

Expanding $T_{\mu\nu}^{CT}$ at large $r$, the first two terms give (\ref{eq:Tgen1-main}) with
\be
b_{CT1}=b_{CT2}=1\,.
\ee
The flow equation in $d$-dimensions for the cylinder is then obtained by writing the extrinsic curvature $K_{\mu\nu}$ as a function of $T_{\mu\nu}+ T^{CT}_{\mu\nu}$ using (\ref{eq:Tct1}),
\be\label{eq:WdWflow}
4\pi G \ell \left((T_{\mu\nu}+T_{\mu\nu}^{CT})^2- \frac{1}{d-1} (T_\mu^\mu+T^{CT\;\mu}_\mu)^2\right)+\frac{\ell}{16\pi G} R^{(d)}+\frac{1}{16\pi G}  \frac{d(d-1)}{\ell}\eta=0\,.
\ee

The more standard version of the flow equation $\tr(T)= \ldots$ is obtained by splitting the $R$-independent counterterm from $T_{\mu\nu}^{CT}$, corresponding to $b_{CT1}$ in (\ref{eq:Tgen1-main}). Writing
\be
T_{\mu\nu}^{CT}= \t T_{\mu\nu}^{CT}+ \frac{d-1}{8\pi G \ell} h_{\mu\nu}
\ee
and defining
\be\label{eq:tTdef}
\t T_{\mu \nu} \equiv T_{\mu\nu}+ \t T_{\mu\nu}^{CT}\,,
\ee
eq. (\ref{eq:WdWflow}) becomes
\be\label{eq:flow2}
\t T^\mu_\mu=4\pi G \ell \left((\t T_{\mu\nu})^2- \frac{1}{d-1} (\t T_\mu^\mu)^2\right)+\frac{\ell}{16\pi G} R^{(d)}+\frac{1}{16\pi G}  \frac{d(d-1)}{\ell}(\eta-1)\,.
\ee

The works \cite{Hartman:2018tkw, Taylor:2018xcy} already laid out the $T^2$ deformation for $d>2$ at the large N level. Here we are adding the generalization to $\eta <0$ (positive 4d bulk cosmological constant) involved in the $T^2+\Lambda_d$ part of the deformation as described in \S\ref{sec-solvable-deformation}, as well as the large N holographic renormalization prescription valid for general $d$ for the cylinder.

Translating to boundary theory language using the dictionary generalizing \S\ref{sec-solvable-deformation}, 
\be\label{eq:translation}
C_d =\frac{1}{4d}\frac{\ell^{d-1}}{G}\;,\;\frac{\lambda}{V^{d/(d-1)}} = \frac{4d G \ell}{V_c^{d/(d-1)}}\,,
\ee
(recall that $V$ is the field theory cylinder volume, while $V_c$ is the volume of the bounding cylinder in the gravity side, which changes along the trajectory)
gives
\be\label{eq-TFE-bdry-var}
\t T^\mu_\mu= \frac{\pi \lambda}{d} :\left(\t T_{\mu\nu} \t T^{\mu\nu}- \frac{1}{d-1} (\t T^\mu_\mu)^2 \right):-\frac{d^2(d-1)}{4\pi} \frac{\eta-1}{\lambda}+ \frac{d}{4\pi} \left(\frac{C_d^2}{\lambda^{d-2}} \right)^{1/d} R^{(d)}\,.
\ee
where $\t T_{\mu\nu}$, defined in (\ref{eq:tTdef}),
includes the contributions to $T_{\mu\nu}$ that arise from counterterms in the boundary action. They can be written in deformed field theory terms using (\ref{eq:translation}) and (\ref{eq:Tct2}).

To connect to the solvable theory defined above in \S\ref{sec-solvable-deformation}, we 
substitute in the special form \eqref{eq-Tab-equilibrium-solvable}, and solving the resulting differential equation for $E(V)$ yields the dressed energy formula
\be\label{eq:dressed-gen}
{\cal E}_n(y)=\frac{d(d-1)}{2\pi y} \left[\sqrt{1+ \Omega^{2/(d-1)} (C_d y)^{2/d}}\Big|_{y^{ (\lceil d-2 \rceil)/d }} \mp \sqrt{\eta+\Omega^{2/(d-1)} (C_d y)^{2/d}-\frac{4\pi}{d(d-1)}y{\cal E}_n(0)}\; \right]
\ee
where $\Omega$ is the volume of the unit radius $(d-2)$--dimensional sphere, $y=\lambda/V^{d/(d-1)}$, $V= \Omega R^{d-1}$, and ${\cal E}= V^{1/(d-1)} E$. As explained before, the first square root is expanded in powers of $y^{1/d}$ up to order $d-2$. For $d=3$, this gives
\eqref{eq-4d-dressed-energies}-\eqref{eq-4d-dressed-energies2}. In the next section, we will make use of both the solvable theory and the large-N general formula \eqref{eq-TFE-bdry-var} in formulating a prescription for the complete dual theory capturing local physics in $dS_4$ with timelike boundary.

Before getting to that, we make two more comments. 
First, one could imagine generalizing the solvable theory defined by \eqref{eq-Tab-equilibrium-solvable} to accurately capture not just the entropically dominant states, but also other types of relatively simple states, such as those with angular momentum. 
This could be obtained from a more general definition (\ref{eq:Hdef}) for the deformed Hamiltonian,
\be
H_\lambda= \sum_E E_\lambda |E \rangle \langle E |
\ee
as the solution to
\bea\label{eq:proposal}
 d \lambda \partial_\lambda (E_\lambda+E_\lambda^{CT})&=& \int\,d^{d-1} x\,\sqrt{-h} \Big[ \frac{\pi \lambda}{d} \left((\langle E | \t T_{\mu\nu}|E \rangle)^2- \frac{1}{d-1}(\langle E | \t T^\mu_{\mu}|E \rangle)^2 \right) \nonumber\\
 &&\qquad \qquad -\frac{d^2(d-1)}{4\pi} \frac{\eta-1}{\lambda}+ \frac{d}{4\pi} \left(\frac{C_d^2}{\lambda^{d-2}} \right)^{1/d} R^{(d)}\Big]\,,
\eea
with the expectation values of the stress tensor being more general than \eqref{eq-Tab-equilibrium-solvable}.  Angular momentum is natural to consider; indeed it easily fits into the $T\bar T+\Lambda_2$ framework in $d=2$ \eqref{eq-2d-pure-gr-energy}.  In $d=3$, in states with nonzero angular momentum ($T^0_\phi\ne 0$), spherical symmetry is broken and the spatial components of the stress energy tensor do not have isotropic pressure, $T^a_b\ne -\delta^a_b dE/dV$.  It would be interesting to incorporate the required generalization into a solvable model that captures more types of states.  But from the gravity side, it is clear that such states are more involved and ultimately require incorporating faithfully the effects of bulk matter treatment along the lines of \cite{Batra:2024kjl} (see \S\ref{sec-local-algorithm}).
Kerr black holes have oblate horizons, so our spherical boundary at the transition point $y=y_u$ will not line up with the boundary of any rotating black hole.  In gravity language, for these states that contain Kerr black holes, as we bring the boundary back out after the uplift $\eta=1\to \eta=\eta_u<0$, we create a domain wall between AdS and dS regions of the bulk spacetime.  Such configurations go beyond the gravitational sector, involving matter fields that interpolate between the negative and positive cosmological constant regions.  

Secondly, let us make a simple estimate to show that bulk quantum effects such as boundary-induced contributions to the stress energy \cite{Deutsch:1978sc} can be included but are subleading to the classical formulas we have written explicitly here. From the structure of the Einstein equations, the boundary effect on the renormalized bulk stress energy $\langle T_{MN}\rangle$ is of order 
\begin{equation}\label{eq-qm-stress-energy}
   \Delta (\frac {g'}{g})^2|_{quantum}\sim  G_N \langle T \rangle \sim \frac{G_N}{\ell_*^2 V_c} \,.
\end{equation}
Here $g'$ represents a proper radial derivative of the metric, while $1/\ell_*$ represents the UV cutoff on this quantum effect near the wall; it is formally divergent at the wall, but will be cut off by the eleven dimensional Planck scale if not before.\footnote{We have not assessed the supersymmetry breaking scale in this setup in detail, but we expect it to be of order the greater of $\eta/\ell$ or $1/\sqrt{V_c}$.}  This back reaction from the quantum stress tensor is to be compared to the classical contribution. From \eqref{eq:End-rescaled}, where the square root term is $\sim \frac{g'}{G_N g}$, this is of order 
\begin{equation}
    (\frac {g'}{g})^2|_{classical} \sim \frac{1}{ \ell_{AdS}^2}
\end{equation}
where we use that at the boundary, $g\sim V_c$.
So the condition that the classical dominates over the quantum is
\begin{equation}
    (\frac{V_c}{\ell_{AdS}^2}) \gg \frac{G_N}{\ell_*^2}\,.
\end{equation} 
Noting that for compactification volume $V_{internal}$, $1/G_N \simeq V_{internal}/\ell_{11}^9$, we see that this condition is satisfied generically along our trajectory; when it fails, semiclassical GR becomes a bad approximation and our theory injects the UV complete details.\footnote{A similar conclusion is reached by comparing the quantum effect with other terms in the Einstein equations. For instance, the contribution from the curvature of the 2-sphere to the equations of motion is of order $1/r_c^2$. This is parametrically bigger than the quantum correction in \eqref{eq-qm-stress-energy}.}

\section{Algorithmic proposal for bulk-local timelike-bounded $dS_4$ holography}\label{sec-local-algorithm}

In this section, we summarize the upgrade to 4 bulk dimensions of the algorithmic prescription \cite{Batra:2024kjl} that builds a deformation that accounts for local bulk matter.  In the present work, the main difference is that the full $T^2$ operator contains stress-energy configurations that are not of the spinless equilibrium form \eqref{eq-Tab-equilibrium-solvable}, including those dual to bulk gravitons as well as angular momentum and shear modes.  This we treat analogously to the treatment of matter in the $dS_3$ case \cite{Batra:2024kjl}. 

In this approach, we first use an infinitesimal stint $(0<y<y_T \ll 1)$ of the solvable $(T^2)_s$ deformation defined in \S \ref{sec-solvable-deformation} as a regulator,  keeping only the real energies in the dressed Hilbert space at $y=y_T$ to produce a unitary finite quantum system.  This deformation introduces some violation of locality, but by choosing the cutoff scale
\begin{equation}
    E_T = \frac{{\cal E}_T}{V^{1/3}} \simeq \frac{3}{2\pi y_T} ~~\leftrightarrow ~~M_T\simeq \frac{R_{cT}^3}{2\ell_{AdS}^2 G}
\end{equation}
sufficiently large (with sufficiently small, but nonzero, $y_T$), we can ensure that there are no violations of locality within the regime of applicability of semiclassical GR + EFT. As described in \S\ref{sec:intro}, we also wish to keep track of any degrees of freedom that will be important in the uplift and which will be light in the dS phase following that step.  We can choose $y_T$ sufficiently small to accommodate this sector.

We then proceed step by step, updating the Hamiltonian at each step by adding renormalized double trace operators required to adjust the boundary conditions to match those of a finite timelike boundary. As described in the previous sections, one can then continue the deformation by adding well-defined regularized versions of the double-trace deformations required for local graviton boundary conditions that had been derived at the large-N level in \cite{Hartman:2018tkw}, incorporating an uplift from AdS to dS at an appropriate point within the deformation. The procedure is summarized in Fig. \ref{fig:setup}.


As discussed above, the status of timelike boundaries in four dimensional GR is still an open research topic, but the special case of a spherical spatial boundary with Dirichlet conditions has not been excluded by any existing analyses, in fact surviving a nontrivial test that flat boundaries fail \cite{Anninos:2023epi}.  Here we will extend this to argue that the quantum theory we construct has no room for non-unique evolution, while having the capacity to match controlled bulk gravity quantities.

For this special Dirichlet case, the appropriate additions follow from the trace flow equation derived in \S\ref{sec:def} (following \cite{Hartman:2018tkw}), but replacing the large-N normal ordered $: T^2:$ operator in \eqref{eq-TFE-bdry-var} by a finite renormalized operator valid at finite N.  Taking into account that
\be
\int h^{\mu\nu} T_{\mu\nu}= R \partial_R S=-d \lambda \partial_\lambda S = d \lambda\partial_\lambda \int dt H\,,
\ee
we write for the parts of the deformation that change $y=\lambda/V^{3/2}$ (i.e. aside from the uplift process)
\be\label{eq:Hdef}
d \lambda \partial_\lambda (H+H_{CT})= \int d^{d-1}x\,\sqrt{-h} \left[ \frac{\pi \lambda}{d} \left(\t T_{\mu\nu} \t T^{\mu\nu}- \frac{1}{d-1} (\t T^\mu_\mu)^2 \right)_r-\frac{d^2(d-1)}{4\pi} \frac{\eta-1}{\lambda}+ \frac{d}{4\pi} \left(\frac{C_d^2}{\lambda^{d-2}} \right)^{1/d} R^{(d)}\right]\,.
\ee
plus matter contributions which are similar to those in \cite{Batra:2024kjl}.  
The composite operator $(T^2)_r$  here is not restricted to the form \eqref{eq-Tab-equilibrium-solvable} 
that entered into the solvable component of the deformation above.  Matter fields entail similar multitrace deformations \cite{Hartman:2018tkw, Batra:2024kjl}, which we can schematically denote $(OO)_r$ in terms of the operator $O$ dual to bulk fields. We will treat the uplift process separately below in \S\ref{sec:uplift}.  In the unregularized theory, one would have difficulty defining the theory obtained by adding these irrelevant operators because of uncontrolled short-distance singularities.  But starting from our large but finite theory, and working step by step along the deformation in theory space, 
we can use the vast Hilbert space to construct well-defined  renormalized multitrace operators $(T^2)_r+(OO)_r+\dots$.    These operators are
defined to match the $EFT + GR + \dots$ quantities that are under control on the gravity side (with the $\dots$ here representing select UV sectors which will be involved in the uplift and $dS$ phases \cite{DeLuca:2021pej}).

At first glance, it may seem like this matching is impossible because of the nominally infinite Hilbert space of the large-N theory, which has a continuum of classical solutions.  One can have black holes with a continuum of masses $M$, as well as bulk gravitons and other quantum field excitations with a continuum of amplitudes.    

However, this continuum is not really reliable; it does not exist at finite $N$.  In GR + EFT, one cannot distinguish between sub-Planckian black hole area differences, corresponding to finite resolution among masses $M$.  The resolution of GR + EFT is limited to $\Delta A \gg G_N$, with $A \sim M G_N^2$. The corresponding resolution of black hole mass differences is $\Delta M \gg 1/\sqrt{G_N}$.  Bulk EFT including graviton excitations form black holes or get absorbed in existing horizons before building up an entropy competitive with the horizon entropy.  Once horizons form, $GR + EFT +\dots$ (including stringy uplift degrees of freedom) does not resolve their microstates.  In contrast, already at the level of the solvable theory in \S\ref{sec-solvable-deformation}, the finite boundary theory contains a dense spectrum of microstates corresponding to each black hole and to the cosmic horizon.  Altogether, there are fewer resolvable states in $GR + EFT+\dots$ than the $\exp(S)$ states that we have in our boundary theory.  
As a result, we can define the operators we need by requiring their matrix elements in our deformed CFT to match the matrix elements within the limited set of states that are resolvable in $GR + EFT+\dots$:
\begin{equation}\label{eq-match-matrix-elements}
    \langle n|_{def-CFT} (T^2)_r | m \rangle_{def-CFT} \equiv \langle n|_{GR+EFT+\dots} (T^2)_r | m \rangle_{GR+EFT+\dots} 
\end{equation}
where $|n\rangle_{GR+EFT+\dots}$ is the state in the resolvable spectrum of $GR + EFT +\dots$  with energy that is closest to that of $|n\rangle_{def-CFT}$ in the deformed-CFT spectrum.  

As a special case of observables to match, we can require our theory to match correlators available on the gravity side, e.g. we define $(T^2(x))_r$ to match
\begin{equation}\label{eq-correlators-tune}
    \langle  ({T}^2(x))_r {O}(y_1)\dots{O}(y_m) \rangle 
\end{equation}
for all operators $O$, down to inter-operators distances sufficiently above the Planck length $G_N^{1/2}$ to be under low energy control.  Again, the ability to tune this follows from the limited resolution of $GR+ EFT+\dots$ along with the vast state space available in a quantum system such as ours.  With Hilbert space dimension $e^S$, the theory has $e^{2S}$ matrix elements available to tune, compared to a more modest number of correlators that are under control in $GR + EFT+\dots$.  

Let us make a rough estimate of this, making use of the fact that
dynamics of the theory (determined by the Hamiltonian) is known to good approximation at a given step, and it can be used to evolve operators to the same time starting from unequal time correlators.  The points in \eqref{eq-correlators-tune} can therefore be taken at a given time, where they commute and there is no operator ordering ambiguity.  
The maximum number of such correlators to match can be estimated as follows.  Suppose we have $k-1$ operators (including matter fields and the components of the stress-energy tensor which correspond to bulk gravitons) that we need to keep track of and match in the effective theory.  In the case of the largest possible boundary,  correlators \eqref{eq-correlators-tune} are constructed by filling each of at most $S$ points with one of these operators or with the identity if no operator is inserted at a given point.  In order to have control over the correlator in GR + EFT, the number of points filled with a nontrivial operator should in fact be $\ll S$, i.e. the number $n_1$ filled with the identity operator should be of order $S$.  The number of ways of assigning the S (distinguishable) points to the identity and $k-1$ operator choices 
is $\frac{S!}{n_1! n_{O_1}!\dots n_{O_{k-1}}!}$. Since $n_1$ is of order $S$, this number is a power law in $S$, i.e. $\ll e^S$.  
Altogether the required tunings for this set of observables are at most a power law in $S$, which is much less than the $e^{2S}$ capacity of the large but finite quantum system.

For a small boundary cylinder near the pole, there are not many operators that fit at a given time while keeping them separated by $\gg \sqrt{G_N}$.  
For this system, a set of correlators that probe deeper into the bulk are those with operators separated in time; it is interesting to consider these directly (without Hamiltonian evolution). For longer times they detect effects deeper into the bulk, toward the horizon.  Restricting to timescales of order $e^S$, related to the metastability of the uplifted de Sitter \cite{DeLuca:2021pej}, one can probe very close to the horizon.  Again, with $e^S \ll e^{2 S}$, the number of correlators to match here is parametrically smaller than the number of matrix elements we have to tune.  

Again, our argument that this tuning is possible is based on the enormous count of tunable parameters in our large but finite system.  This counting argument does not make direct use of any additional structure in the problem.  We expect that one could bring in such structures, such as the bulk locality that, as it stands, we simply tune to match, to potentially a priori reduce the number of parameters that must be tuned.  But this is unnecessary for the main purpose of holography, which is to establish a well defined quantum gravity theory for positive cosmological constant, while addressing basic questions about entropy and observables.  In other words, each side of the duality is effective for different quantities. For the state count, the boundary theory is the useful description, whereas for local bulk physics, the gravity side (to which we match by construction) is the effective description.

The Hamiltonian at the next step is then obtained by adding to the Hamiltonian at the current step these regulated double-trace operators, according to \eqref{eq-TFE-bdry-var} (with the large-N normal ordering replaced by the finite $(T^2)_r$ operator defined here).
We also apply this prescription to the sector which, during the uplift process described shortly in \S\ref{sec:uplift}, achieves an uplift from AdS to dS at $y=y_u$.  
At the level of the solvable model, this is a process which continuously changes $\eta$ from $\eta=1$ to $\eta=\eta_u<0$, while in the full theory it continuously transforms the EFT spectrum from that of AdS to that of the dS uplift. In the model \cite{DeLuca:2021pej}, this uplift involves topology change, which we expect to be physically possible in string theory, analogously to the transitions involving wrapped strings and branes in similar models \cite{Adams:2005rb}.  Given that assumption, the dS model is accessible from AdS/CFT.  This enables a process within the algorithm that uplifts the boundary condition on the degree of freedom $\Phi_u$ that interpolates between AdS and dS.  All of this is a direct generalization of the algorithm expressed for $d=2$ in \cite{Batra:2024kjl}.    

\subsection{The uplift part of the deformation}\label{sec:uplift}

In this section we will outline the steps in the algorithm which resolve the uplift from negative to positive bulk cosmological constant $\Lambda_4 \propto -\eta/\ell_{AdS}^2$.
At the level of the solvable model, this amounts to changing $\eta$.  In the full algorithm, we use fine tuning to accurately treat the matter sectors which effect this uplift.  These include a scalar $\Phi_u$ which interpolates from an AdS minimum (such as the M2-brane CFT \cite{Maldacena:1997re} $AdS_4\times S^7$ with $F_7$ flux in M theory) to a dS minimum (such as a member of the discretuum in the explicit compactifications \cite{DeLuca:2021pej}).  In the examples \cite{DeLuca:2021pej}, this interpolation involves a topology change; as mentioned earlier this assumes that the M theory configuration space is connected, similarly to stringy topology changing transitions studied in  \cite{Adams:2005rb}, with $\Phi_u$ representing the condensate effecting the topology change analagous to the winding modes in \cite{Adams:2005rb}.

The spectrum of light EFT fields in the bulk changes during the uplift.  In general, the bulk action $S_{bulk}$ contains interactions between $\Phi_u$ and other fields, which we will collectively denote $\Phi_\perp$.  In particular, the potential $V(\Phi_u, \Phi_\perp)$ will contain $\Phi_u$-dependent mass terms for the $\Phi_\perp$ fields, as in the related discussion in \cite{Batra:2024kjl}.  

The uplift consists of changing the boundary condition for $\Phi_u$, denoted $\Phi_{uc}$, from its AdS to its dS value.  In the boundary theory variables, this corresponds to changing the corresponding coupling constant $g_u$ in the theory.  
We wish to derive the analogue of \eqref{eq:Hdef} for this uplift process, determining the effect on the spectrum of this interpolation.  We apply the standard bulk/boundary dictionary as laid out in \cite{Gubser:1998bc, Witten:1998qj} and \cite{Balasubramanian:1999re}. 
Namely, at the level of large-N saddle point solutions and their corresponding energy bands in the boundary theory, we have 
\begin{equation}
\partial_{g_u}\langle n|H | n\rangle_{def-CFT} \equiv 
    \partial_{\Phi_{cu}}\langle n| H |n \rangle_{GR+EFT+\dots} = -\partial_{\Phi_{uc}} S_{bulk}\big[ {\bf \Phi}_{*, n}(\Phi_{uc}) \big] 
\end{equation}
where ${\bf \Phi}_*(\Phi_{uc})$ denotes the $nth$ saddle point solution for the $\Phi_u, \Phi_\perp$ fields, which depends on the coupling $\Phi_{uc}$. As in the discussion above around \eqref{eq-match-matrix-elements}, the fine grained states are matched to the closest energy state among this cruder state space resolvable in $GR+EFT+\dots$, up to negligible differences that can be included without affecting the approximate agreement to the bulk.   That is, $|n\rangle_{GR+EFT+\dots}$ is the state in the resolvable spectrum of $GR + EFT +\dots$  with energy that is closest to that of $|n\rangle_{def-CFT}$ in the deformed-CFT spectrum.
This leads to blocks of nearly degenerate states in our finite Hamiltonian. The total number of states $\exp(S)$ 
contains logarithmic corrections to the entropy.
This includes certain corrections $\propto \log(A/4G_N)$ to the entropy which follows from the symmetry structure in our bounded geometry (as explained in \cite{Anninos:2020hfj, Chandrasekaran:2022cip}).\footnote{We thank Dio Anninos and Albert Law for very helpful discussions of this point.}  Since the symmetry structure is the same in the bounded AdS black hole and the bounded dS cosmic horizon patch, the result for this contribution should automatically match as we proceed through the uplift.  Determining this coefficient would arise from a calculation analogous to those in \cite{Anninos:2020hfj, Chandrasekaran:2022cip} and \cite{Sen:2012dw}.\footnote{It may also be possible to generalize the CFT analysis of \cite{Benjamin:2023qsc}} to our case.
It would be interesting to analyze this point further, in order to determine whether the matching of the full EFT + GR +\dots prediction for the corrected entropy is automatic, including matter contributions as well as gravitational and matter counterterms \cite{Anninos:2020hfj}.  If not, it would be possible to make small adjustments to the state count in our spectrum to match the bulk effective theory prediction.

Similarly to the tuning procedure for $(\Delta T^2)_r + (OO)_r$ required to define $\partial_\lambda \langle H \rangle$, this procedure matches low energy\footnote{compared to the 4d Planck energy density $\sim 1/G_N^2$} $GR + EFT +\dots$ effects.  
A difference that we should address, however, is that for bands of energy levels in which in the gravity-side description the boundary is close to a horizon, the fluctuations away from classical saddles are large as stressed in \cite{Silverstein:2022dfj}.  So are the uncertainties in the formulation of the boundary conditions in GR.    
As in \cite{Batra:2024kjl} and above, we treat this as a feature rather than a bug:  our Hamiltonian, constructed as we just described, is finite and well defined.  It simply defines the full quantum gravity theory in this regime where no approximation to it is known.  

For these bands of energy levels, once we finish the uplift (third step of the process, cf. Fig. \ref{fig:setup}) and then evolve in $y=\lambda/V^{3/2}\sim \ell G_N/V_c^{3/2}$ with $\eta=\eta_u$ as in \eqref{eq:Hdef} to where the boundary is no longer near the horizon, low energy $GR + EFT +\dots$ becomes a good approximation to the bulk physics, and our Hamiltonian continues to match to that while filling in finer details of quantum gravity where needed.

\subsection{Comments on the Dirichlet problem in the quantum theory}

\subsubsection{Role of sources}

One way to compute $n$-point functions is to turn on boundary sources, and we should understand whether the geometric uniqueness obtained for the cylinder survives in their presence. 

Let's focus on the correlators for the stress tensor. We turn on a boundary metric source $h^{(0)}_{mn}(x)$; this source is vanishingly small, since we only use it for computing correlators and then set it to zero. We expand the bulk gravitational metric to linear order around the background solution $\bar g$,
\be
g_{\mu\nu}(r, x) = \bar g_{\mu\nu}(r,x) + h_{\mu\nu}(r,x)\,.
\ee
The linearized perturbation (in the absence of matter) satisfies the linearized Einstein equations with inhomogeneous Dirichlet boundary condition
\be
(L h)_{\mu\nu}=0\;,\; h_{mn}(r_c, x) = h^{(0)}_{mn}(x)\,.
\ee
Here $m, n$ are indices parallel to the boundary $r=R_c$.

Computing the gravitational on-shell effective action to quadratic order in $h^{(0)}$ is sufficient for obtaining the stress tensor 2-point function. Higher point functions can be obtained as usual via a diagrammatic expansion like Witten diagrams.

The linearized problem with homogeneous Dirichlet boundary condition $h_{mn}(r_c, x)=0$ on the cylinder plus initial conditions has a unique solution \cite{Anninos:2023epi}. In particular, if the initial conditions are trivial, the only solution is the vanishing solution. This implies that the inhomogeneous problem also has a unique solution. Indeed, if there were different solutions, we could consider their difference; this is a solution to the trivial boundary+initial value problem and hence must vanish. So we cannot have different solutions for the inhomogeneous problem. 

\subsubsection{Quantum uniqueness}

We close with an argument that the quantum theory we have defined in fact can provide a novel way to avoid the classical issue of geometric non-uniqueness.

The phenomenon of geometric non-uniqueness in gravity involves multiple future evolutions from given initial data within the bounded patch.  Since our system has a time translation symmetry, these multiple solutions must share the same Brown-York energy.  But at the quantum level, 
our quantum theory generically has a non-degenerate Hilbert space; when degeneracies appear we may be able to break them slightly consistently with all of our requirements.  For such a Hamiltonian, it is impossible to have multiple evolutions of a given initial energy eigenstate.  

In any case, as we have stressed above, there is no known classical violation of geometric uniqueness (or any other consistency condition) with our Dirichlet condition on a cylinder \cite{Anninos:2023epi} in the first place.

\section{Summary and future directions}

To summarize, this note has shown the following.   

We assume a connected path from $AdS_4$ to $dS_4$.  An example is the topology change from the M2-brane theory dual to $AdS_4\times S^7$ with 7-form flux to its uplift \cite{DeLuca:2021pej}. We expect that such a process is likely to exist analogously to the process studied in \cite{Adams:2005rb}; there is no symmetry forbidding it.  Based on the nontrivial results supporting well-posedness of the cylindrical Dirichlet boundary, as discussed in \cite{Anninos:2023epi} and anticipated at the Euclidean level in  \cite{Witten:2018lgb}.\footnote{We thank M. Anderson, D. Anninos, and D. Galante for very helpful discussions of the  
space of special Dirichlet conditions which do not suffer from the ill-posedness found in the generic case.} we work with this boundary condition.    

Given that, there is a continuous deformation from $AdS_4/CFT_3$ to timelike-bounded $dS_4$ holography, summarized in Fig. \ref{fig:setup}, which matches 

$\bullet$ the horizon entropy as a microstate count 

$\bullet$ the radial geometry of the $dS_4$ cosmic horizon patch and the pole patch, via its correspondence to the Brown-York energy as a function of the proper size $V_c$ of the boundary

$\bullet$ bulk excitations such as gravitons, small black holes, and rotating black holes.  

This result is obtained by breaking the problem into two parts, one of which is analytically solvable and accounts for the leading contribution to the physics of the most entropic energy bands, with the other defining residual effects indirectly via an update algorithm moving step by step in theory space.
The radial geometry and Gibbons-Hawking entropy are captured by a solvable deformation by $(T^2)_s+\Lambda_3$ which is directly analogous to the $T\bar T+\Lambda_2$ deformation in \cite{Gorbenko:2018oov, Coleman:2021nor}.  The algorithmic definition of the full renormalized $T^2$ theory defines the rest of the $T^2 = (T^2)_s + (\Delta T^2)_r$ operator by a tuning prescription matching it to the low energy\footnote{compared to the 4d Planck scale} observables of the bulk theory. The latter are underconstraining compared to the capacity of our finite but enormous quantum Hamiltonian, with its $e^{S}$ states.

The main point of this work is that the deformation method extends to the 4d bulk.  In summary, the solvable $(T^2)_s +\Lambda_3$ deformation captures the $dS_4$ entropy and the behavior of the most entropic energy band, while residual effects are treated accurately with a brute force algorithm.  

Future directions generally include developing further the theory of timelike boundaries from both the macroscopic and microscopic perspectives, including generalizing the entropy corrections \cite{Anninos:2020hfj} to the timelike-bounded case. It will be interesting to derive the phenomenology of this type of topology.  A natural question holographically is the analogous deformation of AdS/CFT for conformal boundary conditions \cite{Bredberg:2011xw, Anninos:2011zn, Anderson:2006lqb, An:2021fcq, Anninos:2023epi, Anninos:2024wpy, Banihashemi:2024yye} and generalizations \cite{Liu:2024ymn}.  Another future direction is to connect our algorithmically built theory to complementary approaches to more direct specifications of a microscopic dual without using a deformation, e.g. via trading flux for branes \cite{Silverstein:2003jp}.

\noindent{\bf Acknowledgements}

We are grateful to G. Batra, G. B. De Luca, and S. Yang for  many useful discussions on de Sitter holography and collaboration on the algorithm \cite{Batra:2024kjl} which underlies our method here, as well as V. Shyam for many discussions of the higher dimensional case, R. Soni for very useful discussions probing the algorithmic approach, and D. Anninos for many insights and explanations on recent developments relevant for this work. We are also very grateful to M. Anderson, B. Banihashemi, F. Denef, L. Freidel, D. Galante, A. Law, C. Maneerat, D. Marolf, J. Santos, and E. Shaghoulian for very useful discussions. E.S. is grateful to the organizers and participants of the August 2024 workshop `Timelike Boundaries in theories of gravity', in Morelia, Mexico, the March 2024 workshop `Quantum Universe' in Chicago, USA and the April meeting `Spacetime and String Theory' at the KITP.   
This research is supported by a Simons Investigator award and National Science Foundation grant PHY-2310429. GT is supported by
CONICET (PIP grant 11220200101008CO),  CNEA, and Instituto Balseiro.

\appendix

\bibliographystyle{JHEP}
\bibliography{refs.bib}
\end{document}